\documentclass[a4paper,11pt]{article}
\usepackage{amsfonts,amsthm,amsmath,amssymb,graphicx,hyperref,color}

\usepackage[vcentermath]{youngtab}

\def\corr#1{\left\langle #1 \right\rangle}
\newcommand{\tr}{\operatorname{tr}}

\newcommand{\Sym}{\operatorname{Sym}}

\def\Dim{\textrm{Dim}}

\def\Sym{\textrm{Sym}}

\def\a{\alpha}
\def\b{\beta}

\def\e{\epsilon}

\def\m{\mu}

\def\n{\nu}

\def\r{\rho}

\def\s{\sigma}

\def\l{\lambda}
\def\L{\Lambda}

\def\D{\Delta}

\def\bI{\mathbb{I}}

\newcommand{\cN}{\mathcal N}
\newcommand{\cO}{\mathcal O}

\newcommand{\be}{\begin{equation}}
\newcommand{\bea}{\begin{eqnarray}}

\newcommand{\ee}{\end{equation}}
\newcommand{\eea}{\end{eqnarray}}

\newcommand{\nn}{\nonumber}

\def\tyng(#1){\hbox{\tiny$\yng(#1)$}\,}
\def\tyoung(#1){\hbox{\tiny$\young(#1)$}}
\newcommand{\scdots}{\scriptstyle{\cdots}}

\begin{document}

\rightline{QMUL-PH-08-02}

\vspace{4truecm}

\centerline{\LARGE \bf Permutations and the Loop}

\vspace{1truecm}

\centerline{{\large \bf T.W. Brown${}^\star$}}

\vspace{.4cm}
\centerline{{\it  Centre for Research in String Theory, Department of Physics}}
\centerline{{ \it Queen Mary, University of London}}
 \centerline{{\it Mile End Road, London E1 4NS, UK}}

\vspace{1.5truecm}

\thispagestyle{empty}

\centerline{\bf ABSTRACT}

\vspace{.5truecm}

\noindent We consider the one-loop two-point function for multi-trace
operators in the $U(2)$ sector of $\cN=4$ supersymmetric Yang-Mills at
finite $N$.  We derive an expression for it in terms of $U(N)$ and
$S_{n+1}$ group theory data, where $n$ is the length of the operators.
The Clebsch-Gordan operators constructed in \cite{Brown:2007xh}, which
are diagonal at tree level, only mix at one loop if you can reach the
same $(n+1)$-box Young diagram by adding a single box to each of the
$n$-box Young diagrams of their $U(N)$ representations (which organise
their multi-trace structure).  Similar results are expected for higher
loops and for other sectors of the global symmetry group.

\vfill
\noindent{\it ${}^\star$t.w.brown@qmul.ac.uk}

\newpage

\tableofcontents

\setcounter{footnote}{0}


\section{introduction}

$\cN=4$ supersymmetric Yang-Mills has three complex scalars
transforming in the adjoint representation of the gauge group $U(N)$.
We focus on operators built out of two of the complex scalars, $X$ and
$Y$, which is a $U(2) \subset SU(4) \subset PSU(2,2|4)$ subsector of
the full global symmetry group of the theory.  Their basic correlators
are given in terms of their $U(N)$ fundamental and antifundamental
indices
\begin{align}
  \corr{ X^{\dagger}{}^i_j(x)\; X^k_l(0)}  =   \corr{ Y^{\dagger}{}^i_j(x)\;
    Y^k_l(0)} & = \frac{1}{x^2}\;\delta^i_l\, \delta^k_j\nn \\
  \corr{ X^{\dagger}{}^i_j(x)\; Y^k_l(0)} & = 0 \label{eq:basic}
\end{align}
From here onwards we will drop the spacetime dependence of the
correlators and focus on the combinatorial parts. We will use a
convention whereby $\corr{ \cdots}$ means the tree-level correlator
where we Wick contract with \eqref{eq:basic}.

We can build gauge-invariant operators by taking traces such as
$\tr(Y)\tr(XYX)$ or $\tr(XXYY)$.  These can be written by letting
permutations act on the gauge indices
\begin{equation}
  \tr(Y)\tr(XYX) = X^{i_1}_{i_4} X^{i_2}_{i_1} Y^{i_3}_{i_3}
  Y^{i_4}_{i_2} =  X^{i_1}_{i_{\a(1)}}  X^{i_2}_{i_{\a(2)}}
  Y^{i_3}_{i_{\a(3)}}  Y^{i_4}_{i_{\a(4)}} \equiv \tr(\a\,XXYY)
\end{equation}
Here $\a = (142)$ is an element of the symmetric group $S_4$ of
permutations of four objects.

In this paper we derive an expression for the one-loop two-point
function of these operators in terms of this group-theoretic language.
In essence all this requires is that we follow permutations and
double-line index loops \cite{'t Hooft:1973jz} carefully.  We make
extensive use of the representation theory methods developed for the
$U(1)$ sector in \cite{Corley:2001zk} and the diagrammatic techniques
introduced in \cite{Corley:2002mj}.

At tree level the correlator in terms of permutations is \cite{Brown:2007xh}
\begin{equation}
  \corr{ \tr(\a_2\, X^{\dagger \m} Y^{\dagger \n})\;  \tr(\a_1\, X^\m
    Y^\n) } = \frac{1}{\m!\n!}\sum_{\s,\tau \in S_{\m} \times S_{\n}
  }\;\sum_{T\,\vdash\, n}\chi_T(\s^{-1}
  \a_1 \s\;\tau^{-1} \a_2 \tau ) \Dim T \label{eq:treeperm}
\end{equation}
Here $X^\m$ just means $\m$ copies of $X$ ($\mu$ is a power not an
index) and similarly for $Y$.  $S_\m \times S_\n$ is the subgroup of
the symmetric group $S_{\m+\n}$ that doesn't mix the first $\m$ items
with the last $\n$, reflecting the fact that $X$ does not mix with $Y$
when we Wick contract with \eqref{eq:basic}\footnote{this expression
  for the tree level correlator is a tad redundant because we can
  absorb the $\tau$ sum into the $\s$ sum; we have written it like
  this to emphasis the comparison with the one-loop case}.  We sum
over all $n\equiv \m+\n$ box Young diagrams $T$ with at most $N$ rows,
each of which labels an irreducible representation both of $S_n$ and
of $U(N)$.  This Schur-Weyl duality of the irreducible representations
of $S_n$ and $U(N)$ follows because they have a commuting action on
$V_N^{\otimes n}$ where $V_N$ is the fundamental representation space
of $U(N)$.  $\chi_T$ is an $S_n$ character and $\Dim T$ is the
dimension of the $U(N)$ representation.  Because $T$ has $n$ boxes its
leading large $N$ behaviour is $\Dim T \sim k N^n$ (see identity
\eqref{eq:dimR}).

In \cite{Brown:2007xh} a basis $\cO[ \L, \m,\n,\b; R;\tau ]$ was found
that diagonalises this tree-level two-point function. $[ \L,
  \m,\n,\b]$ labels the $U(2)$ representation and state while $R$
labels the $U(N)$ representation which organises the multi-trace
structure\footnote{the operator as a whole is a $U(N)$ singlet since
  it is gauge-invariant}.
\begin{figure}[t]
\begin{center}
\includegraphics[trim=0 -10 0 -10]{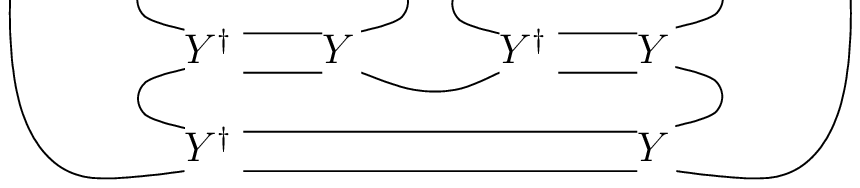}
\caption{a planar one-loop diagram for a part of the two-point
  function between $\tr(XXYY)$ and $\tr(X^\dagger X^\dagger Y^\dagger
  Y^\dagger)$ with the $\tr(YXX^\dagger Y^\dagger)$ effective vertex;
  note this leading $N^{4+1}$ behaviour}\label{fig:intro}
\end{center}
\end{figure}

At one loop we get corrections from the self-energy, the scalar
four-point vertex and the exchange of a gluon.  Cancellations among
these corrections mean that the one-loop correlator is given by an
effective vertex \cite{Constable:2002hw}\cite{Beisert:2002bb}
\begin{equation}
\corr{ \tr(\a_2\, X^{\dagger \m}
  Y^{\dagger \n}) :\tr([X,Y][X^\dagger,Y^\dagger] ): \tr(\a_1\, X^\m
    Y^\n)} \label{eq:normalorder}
\end{equation}
For convenience we have dropped a $ - \frac{g_{\textrm{YM}}^2}{8\pi}$
prefactor and the spacetime dependence $\log(x\L)^{-2}/x^{2n}$ for
some cutoff $\L$.  The expression betwen colons $: :$ is
normal-ordered so that no contractions within the colons is allowed.
In Sections \ref{sec:dil} and \ref{sec:oneloop} we derive an
expression for this one-loop correlator in terms of permutations
\begin{equation}
 \frac{1}{(\m-1)!}  \frac{1}{(\n-1)!}\sum_{\s,\tau \in S_\m \times
     S_\n}\sum_{\r_1,\r_2\in S_{n+1}} h(\r_1,\r_2) \sum_{T\,\vdash\,n+1}\chi_T(\r_1\; \s^{-1}
  \a_1 \s \; \r_2 \;\tau^{-1} \a_2
  \tau )\Dim T
\end{equation}
Compare this with \eqref{eq:treeperm}. Now $T$ has $n+1$ boxes and
$\chi_T$ is a character of $S_{n+1}$.  For large $N$ the leading
behaviour is $\Dim T \sim kN^{n+1}$, which is what we expect for the
one-loop result (see for example Figure \ref{fig:intro}).
$h(\r_1,\r_2)$ only takes non-zero values on a few permutations of the
$\m$, $n$ and $n+1$ indices (it is given in full in equation
\eqref{eq:fullh}); it encodes the commutators in
\eqref{eq:normalorder}.

We also derive a similar expression for the one-loop dilatation
operator.

We find that the Clebsch-Gordan basis $\cO[ \L, \m,\n,\b; R;\tau ]$
has constrained mixing at one loop.  If two operators are in the same
$U(2)$ representation and state, then if their $U(N)$ representations
$R_1$ and $R_2$ are different they only mix if we can add a box to
each Young diagram to get the same $U(N)$ representation with $n+1$
boxes $T$. For example $R_1 = \tyng(2,2)$ and $R_2 = \tyng(3,1)$ mix
because we can get them both by knocking a single box off $T =
\tyng(3,2)$.  In other words, when we restrict the representation $T$
of $S_{n+1}$ to its $S_n$ subgroup, $R_1$ and $R_2$ must both appear
in the reduction.  This mixing is analysed in Section
\ref{sec:mixing}.  A detailed look at the $U(2)$ representation $\L =
\tyng(2,2)$ operators is given in Appendix Section \ref{sec:example}.

Extensions to higher loops and the rest of the global symmetry are
discussed in Section \ref{sec:extensions}.

Appendix \ref{sec:convandform} covers some group theory conventions
and formulae; Appendix \ref{sec:diagrammatics} briskly introduces the
diagrammatic formalism we use; Appendix \ref{sec:symmetricgroup}
revises the construction of the representing matrices for the
symmetric group.

\section{the dilatation operator}\label{sec:dil}

Given that $\corr{ X^{\dagger}{}^i_j\; X^k_l} = \tilde X^i_j \;X^k_l =
\delta^i_l \delta^k_j$ where $\tilde X^i_j = \frac{d }{dX^j_i}$ we can
get the one-loop correlator by first acting on $\tr(\a_1\, X^\m Y^\n)$
with the one-loop dilatation operator
\cite{Beisert:2002bb}\cite{Gross:2002mh}\cite{Janik:2002bd}\cite{Beisert:2002ff}
\begin{equation}
 \D^{(1)} =  \tr([X,Y][\tilde X,\tilde Y] )
\end{equation}

As a warm-up consider the action of $\tilde X^{a}_{b}$ on
\begin{equation}
  X^{i_{1}}_{j_{1}} \cdots   X^{i_{n}}_{j_{n}}
\end{equation}
By the product rule we get
\begin{equation}
  \left(\delta^{a}_{j_1}  \delta^{i_1}_{b} \right)\; X^{i_{2}}_{j_{2}} \cdots
  X^{i_{n}}_{j_{n}}\;\; + \;\;  X^{i_{1}}_{j_{1}}\; \left(\delta^{a}_{j_2}
  \delta^{i_2}_{b} \right)\;  X^{i_{3}}_{j_{3}} \cdots
  X^{i_{n}}_{j_{n}}\;\; + \;\;\cdots
\end{equation}
To write this down in terms of permutations we shuffle around the
$\delta$'s with $\s\in S_n$ so that the derivative only ever acts on
the final index
\begin{equation}
 \frac{1}{(n-1)!} \sum_{\s \in S_n } \left(\delta^{a}_{j_{\s(n)}}  \delta^{i_{\s(n)}}_{b} \right)\; X^{i_{\s(1)}}_{j_{\s(1)}} \cdots
  X^{i_{\s(n-1)}}_{j_{\s(n-1)}} \label{eq:yak}
\end{equation}
We divide by $(n-1)!$ because summing over all of $S_n$ is
redundant\footnote{it would be more economical to sum over $\s \in
  \Sym(n)$, the symmetry group of an $n$-cycle, in which case we would
  not have to divide by $(n-1)!$, but this is not necessary for our
  purposes}.

It is a small step now to the action of  $\tilde X^{a}_{b}\tilde
Y^c_d$ on 
\begin{equation}
  X^{i_{1}}_{j_{1}} \cdots   X^{i_{\m}}_{j_{\m}}
  Y^{i_{\m+1}}_{j_{\m+1}} \cdots   Y^{i_{\m+\n}}_{j_{\m+\n}} \label{eq:covar}
\end{equation}
We get
\begin{equation}
 \frac{1}{(\m-1)!}  \frac{1}{(\n-1)!}\sum_{\s \in S_\m \times S_\n} \left(\delta^{a}_{j_{\s(\m)}}  \delta^{i_{\s(\m)}}_{b}
  \right) \left(\delta^{c}_{j_{\s(\m+\n)}}
  \delta^{i_{\s(\m+\n)}}_{d} \right)\;\;  X^{i_{\s(1)}}_{j_{\s(1)}} \cdots
  X^{i_{\s(\m-1)}}_{j_{\s(\m-1)}} \;\; Y^{i_{\s(\m+1)}}_{j_{\s(\m+1)}} \cdots
  Y^{i_{\s(\m+\n-1)}}_{j_{\s(\m+\n-1)}} \nn
\end{equation}
Next we relabel indices $i_{\s(k)} \to p_k$ and $j_{\s(k)} \to q_k$
for $k \in \{1, \dots \m-1,\m+1, \dots \m+\n-1\}$.  This amounts to
writing $X^{i_{\s(k)}}_{j_{\s(k)}} = \delta^{i_{\s(k)}}_{p_k}
\delta^{q_k}_{j_{\s(k)}} X^{p_k}_{q_k}$, which is just a book-keeping
exercise.\footnote{we advise the reader to glance over Appendix
  \ref{sec:diagrammatics} for the delta function and diagrammatic
  techniques used here}
\begin{align}
& \frac{1}{(\m-1)!}  \frac{1}{(\n-1)!}\sum_{\s \in S_\m \times S_\n} \left(\delta^{a}_{j_{\s(\m)}}  \delta^{i_{\s(\m)}}_{b}
  \right) \left(\delta^{c}_{j_{\s(\m+\n)}}
  \delta^{i_{\s(\m+\n)}}_{d} \right) \nn \\
&\quad\quad\delta^{i_{\s(1)}}_{p_1} \cdots
  \delta^{i_{\s(\m-1)}}_{p_{\m-1}} \;\; \delta^{i_{\s(\m+1)}}_{p_{\m+1}} \cdots
  \delta^{i_{\s(\m+\n-1)}}_{p_{\m+\n-1}} \;\;\;
\delta^{q_1}_{j_{\s(1)}} \cdots
  \delta^{q_{\m-1}}_{j_{\s(\m-1)}} \;\; \delta^{q_{\m+1}}_{j_{\s(\m+1)}} \cdots
  \delta^{q_{\m+\n-1}}_{j_{\s(\m+\n-1)}}
\nn\\
&\quad\quad\quad\quad X^{p_1}_{q_1} \cdots X^{p_{\m-1}}_{q_{\m-1}}
\;\;Y^{p_{\m+1}}_{q_{\m+1}} \cdots Y^{p_{\m+\n-1}}_{q_{\m+\n-1}}
\end{align}
\begin{figure}[t]
\begin{center}
\includegraphics[trim=0 -40 0 -35]{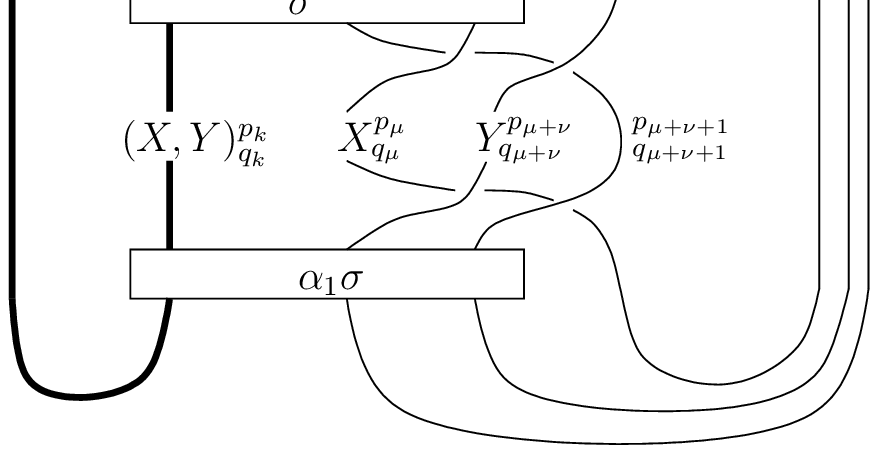}
\caption{the first term $\tr(X Y \tilde X\tilde Y)$ of the one-loop
  dilatation operator acting on $\tr(\a_1 \, X^{ \m} Y^{ \n})$; $k$
  labels the indices in $ \{1, \dots \m-1,\m+1, \dots \m+\n-1\}$ and
  these delta function strands are grouped together into a single
  thick strand; the $\mu$, $\mu+\n$ and $\m+\n+1$ strands are drawn
  separately}\label{fig:firstdil}
\end{center}
\end{figure}Now let's contract some indices.  We're not interested in the
gauge-covariant operator \eqref{eq:covar}; we'd like to know about
$\tr(\a_1 \, X^{ \m}  Y^{ \n})$, which means setting
$j_m = i_{\a_1(m)}$.  Also we need to contract the indices of the
dilatation operator  $\tr([X, Y][\tilde X,\tilde Y])$
\begin{align}
&  \tr(X Y \tilde X\tilde Y) -
  \tr(Y X \tilde X\tilde Y) - \tr(X Y\tilde  Y\tilde X) + \tr(Y
  X\tilde  Y\tilde X) \nn \\
   & =  X^{p_\m}_{q_\m}  Y^{p_{\m+\n}}_{q_{\m+\n}}
  \tilde X^{a}_{b} \tilde Y^{c}_{d} \left(
\delta^{q_\m}_{p_{\m+\n}} \delta^{q_{\m+\n}}_{a}
\delta^{b}_{c}   \delta^{d}_{p_\m} - \delta^{q_\m}_{a} \delta^{q_{\m+\n}}_{p_\m}
\delta^{b}_{c}   \delta^{d}_{p_{\m+\n}}
  - \delta^{q_\m}_{p_{\m+\n}} \delta^{q_{\m+\n}}_{c}
\delta^{b}_{p_\m}   \delta^{d}_{a} + \delta^{q_\m}_{c} \delta^{q_{\m+\n}}_{p_\m}
\delta^{b}_{p_{\m+\n}}   \delta^{d}_{a} \right) \label{eq:dilatation}
\end{align}
This all looks frightful, but let's take the first term of the
one-loop dilatation operator and work it out
\begin{align}
  \tr( X Y \tilde X\tilde Y) \left[\tr(\a_1 \, X^{ \m}  Y^{ \n}) \right] =
   \frac{1}{(\m-1)!}  \frac{1}{(\n-1)!}\sum_{\s \in S_\m \times
     S_\n}\delta^{q_{\m+\n}}_{i_{\a_1\s(\m)}}  \;\;
   \delta^{i_{\s(\m)}}_{i_{\a_1\s(\m+\n)}}
  \delta^{i_{\s(\m+\n)}}_{p_\m} \delta^{q_\m}_{p_{\m+\n}} \nn \\
\delta^{i_{\s(1)}}_{p_1} \cdots
  \delta^{i_{\s(\m-1)}}_{p_{\m-1}} \;\; \delta^{i_{\s(\m+1)}}_{p_{\m+1}} \cdots
  \delta^{i_{\s(\m+\n-1)}}_{p_{\m+\n-1}} \quad\quad
\delta^{q_1}_{i_{\a_1\s(1)}} \cdots
  \delta^{q_{\m-1}}_{i_{\a_1\s(\m-1)}} \;\; \delta^{q_{\m+1}}_{i_{\a_1\s(\m+1)}} \cdots
  \delta^{q_{\m+\n-1}}_{i_{\a_1\s(\m+\n-1)}}
\nn\\
X^{p_1}_{q_1} \cdots X^{p_{\m}}_{q_{\m}}
\;\;Y^{p_{\m+1}}_{q_{\m+1}} \cdots Y^{p_{\m+\n}}_{q_{\m+\n}} \label{eq:ghastly}
\end{align}
Although this still looks rather ghastly, we can see some similarities
emerging between the terms from the dilatation operator on the first
line and those on the second line from the Wick contractions.  They
become clear if we introduce an extra index $\m+\n+1$ and split out
the deltas $\delta^{q_\m}_{p_{\m+\n}} =
\delta^{q_\m}_{i_{\m+\n+1}}\delta^{i_{\m+\n+1}}_{p_{\m+\n}}$ and
$\delta^{i_{\s(\m)}}_{i_{\a_1\s(\m+\n)}}
=\delta^{i_{\s(\m)}}_{p_{\m+\n+1}} \delta^{p_{\m+\n+1}}_{q_{\m+\n+1}}
\delta^{q_{\m+\n+1}}_{i_{\a_1\s(\m+\n)}}$. The expression is now more
pleasing
\begin{align}
  \tr( X Y \tilde X\tilde Y) \left[\tr(\a_1 \, X^{ \m}  Y^{ \n}) \right] =
   \frac{1}{(\m-1)!}  \frac{1}{(\n-1)!}\sum_{\s \in S_\m \times
     S_\n}  X^{p_1}_{q_1} \cdots X^{p_{\m}}_{q_{\m}}
\;\;Y^{p_{\m+1}}_{q_{\m+1}} \cdots Y^{p_{\m+\n}}_{q_{\m+\n}}\; \delta^{p_{\m+\n+1}}_{q_{\m+\n+1}} \nn\\
\delta^{i_{\s(1)}}_{p_1} \cdots
  \delta^{i_{\s(\m-1)}}_{p_{\m-1}} \delta^{i_{\s(\m+\n)}}_{p_\m} \delta^{i_{\s(\m+1)}}_{p_{\m+1}} \cdots
  \delta^{i_{\s(\m+\n-1)}}_{p_{\m+\n-1}}\delta^{i_{\m+\n+1}}_{p_{\m+\n}}\delta^{i_{\s(\m)}}_{p_{\m+\n+1}} \nn\\
\delta^{q_1}_{i_{\a_1\s(1)}} \cdots
  \delta^{q_{\m-1}}_{i_{\a_1\s(\m-1)}} \delta^{q_\m}_{i_{\m+\n+1}} \delta^{q_{\m+1}}_{i_{\a_1\s(\m+1)}} \cdots
  \delta^{q_{\m+\n-1}}_{i_{\a_1\s(\m+\n-1)}}\delta^{q_{\m+\n}}_{i_{\a_1\s(\m)}}  \delta^{q_{\m+\n+1}}_{i_{\a_1\s(\m+\n)}}
\end{align}
Introducing the extra index allows us to draw this diagrammatically as
a trace of a series of operations on the strands, see Figure
\ref{fig:firstdil}.  This was not possible with the expression in
\eqref{eq:ghastly}.  Converting the diagram back to a formula we get
\begin{align}
&    \tr( X Y \tilde X\tilde Y) \left[\tr(\a_1 \, X^{ \m}  Y^{ \n})
    \right] \nn\\
& =   \frac{1}{(\m-1)!}  \frac{1}{(\n-1)!}\sum_{\s \in S_\m \times
     S_\n} \tr\left( \,(\m,\m+\n+1,\m+\n)\,\,\s^{-1}\a_1\s\,\, (\m,\m+\n+1,\m+\n)\, X^\m Y^\n \bI_N\right)
\end{align}
$\bI_N$ is a single $U(N)$ identity matrix and $(\m,\m+\n+1,\m+\n)$ is
a 3-cycle permutation in $S_{n+1}$.
\begin{figure}[t]
\begin{center}
\includegraphics[trim=0 -40 0 -35]{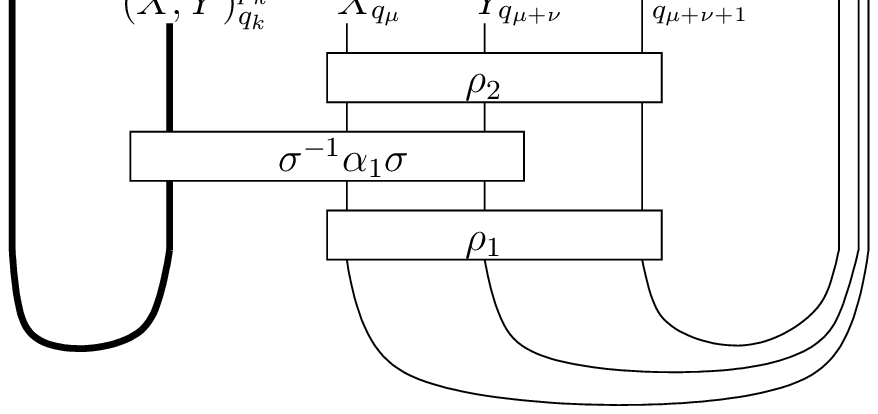}
\caption{the general diagram for any of the four terms of the one-loop
  dilatation operator} \label{fig:dil4}
\end{center}
\end{figure}

If we include the other terms in the one-loop dilatation operator
\eqref{eq:dilatation} then we get
\begin{align}
&  \tr([X, Y][\tilde X,\tilde Y]) \left[\tr(\a_1 \, X^{ \m}  Y^{ \n})
    \right] \nn\\& = \frac{1}{(\m-1)!}  \frac{1}{(\n-1)!}\sum_{\s \in S_\m \times
     S_\n}\sum_{\r_1,\r_2\in S_{n+1}} h(\r_1,\r_2) \tr(\r_1\;
  \s^{-1}\a_1\s\; \r_2\; X^\m Y^\n \bI_N)
\end{align}
See Figure \ref{fig:dil4} for the diagram for general
$\rho_1,\rho_2$. $h$ takes non-zero values on
\begin{align}
  h((\m,n+1,n), (\m,n+1,n)) &= 1 \nn\\
  h((\m,n+1), (n,n+1)) &= -1 \nn\\
  h((n,n+1), (\m,n+1))  &= -1 \nn\\
  h((\m,n,n+1),  (\m,n,n+1)) &= 1  \label{eq:fullh}
\end{align}
We can write this in a more symmetric fashion that better reflects the
commutator structure of the one-loop dilatation operator
\begin{align}
  h(\;(\m,n+1), \;\;(n,n+1)\;) &= -1 \nn\\
  h(\;(\m,n)\;(\m,n+1), \;\; (n,n+1)\;(\m,n)\;) &= 1 \nn\\
  h(\;(\m,n)\;(\m,n+1)\;(\m,n), \;\; (\m,n)\;(n,n+1)\;(\m,n)\;)  &= -1 \nn\\
  h(\;(\m,n+1)\;(\m,n), \;\;  (\m,n)\;(n,n+1)\;) &= 1  \label{eq:fullh2}
\end{align}
We will use this later.

\begin{figure}[t]
\begin{center}
\includegraphics[trim=0 -65 0 -35]{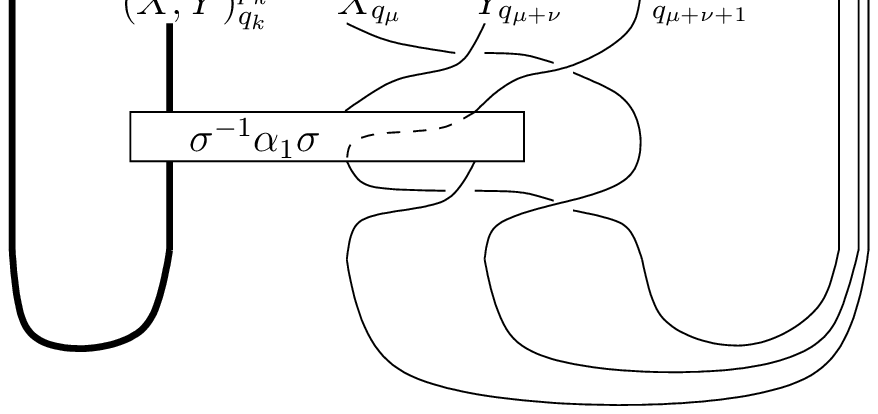}
\caption{an example of how the extra index allows an index loop to
  form, giving an $N$ enhancement} \label{fig:dil3}
\end{center}
\end{figure}

We can see that this extra index gives an enhancement by a factor of
$N$ when a loop forms, see Figure \ref{fig:dil3}.  This happens when
$\s^{-1}\a_1\s$ maps $\m+\n \mapsto \m$ or $\m \mapsto \m+\n$,
i.e. when $X$ and $Y$ are next to each other in a trace $\tr(\cdots XY
\cdots)$.  This is well-studied in the planar context where this
contribution dominates and the model is exactly solvable by the Bethe
Ansatz (see for example
\cite{Minahan:2002ve}\cite{Beisert:2003tq}\cite{Beisert:2003yb}).  In
the non-planar context the trace structure of the operator is still
modified when $\s^{-1}\a_1\s$ does not satisfy this condition, and
traces can split and join (see for example \cite{Bellucci:2004ru}).

\section{one-loop correlator}\label{sec:oneloop}

To get the one-loop correlator we take the tree-level correlator of
$\tr(\a_2 \, X^{\dagger \m} Y^{\dagger \n})$ with the image of
$\tr(\a_1 \, X^{ \m} Y^{ \n})$ under the one-loop dilatation operator
\begin{align}
 & \corr{ \tr(\a_2\, X^{\dagger \m}
  Y^{\dagger \n}) :\tr([X,Y][X^\dagger,Y^\dagger] ): \tr(\a_1\, X^\m
    Y^\n)} \nn \\
& =  \corr{ \tr(\a_2\, X^{\dagger \m}
  Y^{\dagger \n}) \tr([X, Y][\tilde X,\tilde Y]) \left[\tr(\a_1 \, X^{
      \m}  Y^{ \n})\right]} \nn\\
& =  \frac{1}{(\m-1)!}  \frac{1}{(\n-1)!}\sum_{\s \in S_\m \times
     S_\n}\sum_{\r_1,\r_2\in S_{n+1}} h(\r_1,\r_2)\nn\\
 &\quad\quad\corr{ X^{\dagger}{}^{j_1}_{j_{\a_2(1)}} \cdots
    Y^{\dagger}{}^{j_{n}}_{j_{\a_2(n)}}\quad\; X^{i_1}_{i_{\r_1\s^{-1}\a_1\s \r_2(1)}} \cdots
    Y^{i_n}_{i_{\r_1\s^{-1}\a_1\s \r_2(n)}}    \delta^{i_{n+1}}_{i_{\r_1\s^{-1}\a_1\s \r_2(n+1)}}}
\end{align}
Now Wick contract with \eqref{eq:basic}, permuting with $\tau$ for all
the possible combinations
\begin{align}
  & \frac{1}{(\m-1)!}  \frac{1}{(\n-1)!}\sum_{\s,\tau \in S_\m \times
     S_\n}\sum_{\r_1,\r_2\in S_{n+1}} h(\r_1,\r_2)\nn\\
 & \quad  \delta^{j_{\tau(1)}}_{i_{\r_1\s^{-1}\a_1\s \r_2(1)}}\;
 \delta^{i_{1}}_{j_{\a_2\tau(1)}}\;\; \cdots\;\;
 \delta^{j_{\tau(n)}}_{i_{\r_1\s^{-1}\a_1\s \r_2(n)}}\;
 \delta^{i_{n}}_{j_{\a_2\tau(n)}} \quad
 \delta^{i_{n+1}}_{i_{\r_1\s^{-1}\a_1\s \r_2(n+1)}} \nn\\
& =  \frac{1}{(\m-1)!}  \frac{1}{(\n-1)!}\sum_{\s,\tau \in S_\m \times
     S_\n}\sum_{\r_1,\r_2\in S_{n+1}} h(\r_1,\r_2)\;
 \delta^{i_{1}}_{i_{\r_1\s^{-1}\a_1\s \r_2\tau^{-1}\a_2\tau(1)}} \cdots
 \delta^{i_{n+1}}_{i_{\r_1\s^{-1}\a_1\s \r_2\tau^{-1}\a_2\tau(n+1)}}
 \nn\\
&=\frac{1}{(\m-1)!}  \frac{1}{(\n-1)!}\sum_{\s,\tau \in S_\m \times
     S_\n}\sum_{\r_1,\r_2\in S_{n+1}} h(\r_1,\r_2)\; \tr(\r_1\; \s^{-1}
  \a_1 \s \; \r_2 \;\tau^{-1} \a_2
  \tau\;\bI_N^{n+1} )
\end{align}
\begin{figure}[t]
\begin{center}
\includegraphics[trim=0 -40 0 -35]{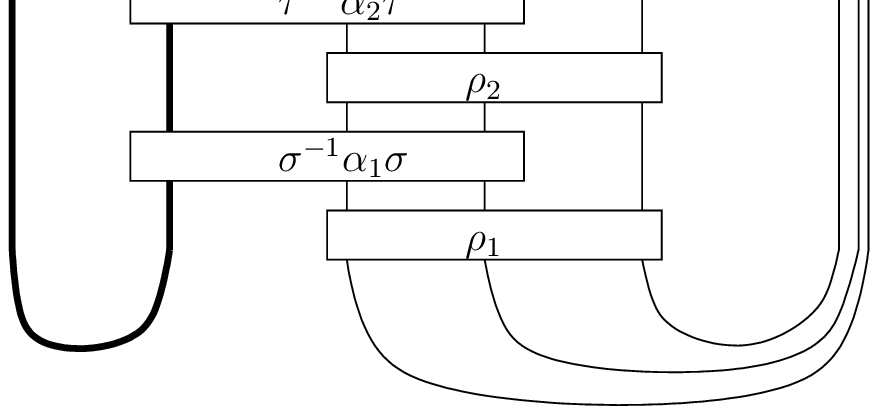}
\caption{one-loop correlator} \label{fig:dil5}
\end{center}
\end{figure}
See Figure \ref{fig:dil5} for the diagrammatic representation of this
trace.  We can expand it in characters of $S_{n+1}$ and dimensions of
$U(N)$ $(n+1)$-box representations
\begin{align}
   & \corr{ \tr(\a_2\, X^{\dagger \m}
  Y^{\dagger \n}) :\tr([X,Y][X^\dagger,Y^\dagger] ): \tr(\a_1\, X^\m
    Y^\n)} \nn \\
& = \frac{1}{(\m-1)!}  \frac{1}{(\n-1)!}\sum_{\s,\tau \in S_\m \times
     S_\n}\sum_{\r_1,\r_2\in S_{n+1}} h(\r_1,\r_2) \sum_{T\,\vdash\,n+1}\chi_T(\r_1\; \s^{-1}
  \a_1 \s \; \r_2 \;\tau^{-1} \a_2
  \tau )\Dim T\label{eq:finalonepoint}
\end{align}

\section{operator mixing}\label{sec:mixing}

Operator mixing between single- and multi-trace operators at one-loop
has been well studied (see for example
\cite{Arutyunov:2000ima}\cite{Arutyunov:2002rs}\cite{Bianchi:2002rw}\cite{Constable:2002vq}\cite{Beisert:2002bb}).
Here we will consider the mixing of a different basis of operators.

In \cite{Brown:2007xh} a complete basis of gauge-invariant operators
was constructed that diagonalises the tree-level correlator for a
theory with $U(M)$ global flavour symmetry and $U(N)$ gauge symmetry.
This Clebsch-Gordan basis tells us how to mesh the $U(2)$ (or more
generally the $U(M)$) representation, which dictates how the operator
transforms under the flavour group, with the $U(N)$ representation,
which controls the multi-trace structure
\begin{align}
  \cO[ \L, \m,\n,\b; R;\tau ]  & \equiv \frac{1}{(n!)^2}\sum_{\a, \s \in S_n} B_{j \b}
  \; S^{\tau ,}{}^{ \L }_{ i }\;{}^{R}_{k }\;{}^{R}_{l}\;\;D_{ij}^\L(\s)
  D_{kl}^R(\a) \tr(\a\s\; X^\m Y^\n \; \s^{-1}) \nn \\
& = \frac{1}{n!}\sum_{\a \in S_n} B_{j \b}
  \; S^{\tau ,}{}^{ \L }_{ j }\;{}^{R}_{p }\;{}^{R}_{q}\;\;
  D_{pq}^R(\a) \tr(\a\; X^\m Y^\n ) \label{eq:simplifiedop}
\end{align}
The equality follows from identity \eqref{Hamer186}.  Here $\L$ labels
the $U(2)$ representation and $[\m,\n,\b]$ labels the state within
$\L$: $\m,\n$ label the number of fields $X,Y$ and $\b \in \{1, \dots
g(\, \hbox{{\tiny$\overbrace{\young(\ \ \scdots\ )}^\mu$}}\, ,
\hbox{{\tiny$\overbrace{\young(\ \ \scdots\ )}^\nu$}}\,;\L) \}$ labels
the semistandard tableau with field content $X^\m$ and
$Y^\n$.\footnote{The Littlewood-Richardson coefficient $g$ counts the
  number of times $\L$ appears in
  $\hbox{{\tiny$\overbrace{\young(\ \ \scdots\ )}^\mu$}}\, \circ
  \hbox{{\tiny$\overbrace{\young(\ \ \scdots\ )}^\nu$}}\,$, where
  $\circ$ is the tensor product for $U(2)$ and the outer product for
  the symmetric group $S_n$. For such tensor products of totally
  symmetric representations, this Littlewood-Richardson coefficient is
  also known as the Kostka number for $\L$ and field content $\m,\n$.
  In the $U(2)$ case this is all a bit trivial because $g(\,
  \hbox{{\tiny$\young(\ \ \scdots\ )$}}\, ,
  \hbox{{\tiny$\young(\ \ \scdots\ )$}}\,;\L)$ is
  either zero or one, but the $\beta$ multiplicity becomes non-trivial
  for $U(M)$ with $M \geq 3$.  $B_{j\beta}$ is the branching
  coefficient for the restriction of $\L$ to the representation
  $\hbox{{\tiny$\overbrace{\young(\ \ \scdots\ )}^\mu$}}\, \circ
  \hbox{{\tiny$\overbrace{\young(\ \ \scdots\ )}^\nu$}}$ of its $S_\m
  \times S_\n$ subgroup.} $R$ labels the $U(N)$ representation, which
dictatess the multi-trace structure of the operator.  $\tau$ labels the
number of times $\L$ appears in the symmetric group tensor product
$R\otimes R$ (also called the inner product).  $S^{\tau ,}{}^{ \L }_{
  j }\;{}^{R}_{p }\;{}^{R}_{q}$ is the $S_n$ Clebsch-Gordan
coefficient for this tensor product\footnote{$S^{\tau ,}{}^{ \L }_{ j
  }\;{}^{R}_{p }\;{}^{R}_{q}$ for $S_n$ is exactly analogous to the
  $3j$-symbol for $SU(2)$, which is just an expression of the
  Clebsch-Gordan coefficients we know and love}.  From the unitary
group perspective $S$ blends the global symmetry $U(2)$ with the gauge
symmetry $U(N)$.  $D^R_{pq}(\a)$ is the real orthogonal
Young-Yamanouchi $d_R \times d_R$ matrix for the representation $R$ of
the symmetry group $S_n$.  It is constructed in Chapter 7 of Hamermesh
\cite{hamermesh} following the presentation by Yamanouchi
\cite{yamanouchi}.  All of these factors are explained in detail in
\cite{Brown:2007xh}.

At tree level these operators are diagonal
\begin{equation}
    \corr{\cO^\dagger[ \L_2, \m_2,\n_2,\b_2; R_2;\tau_2 ]\;
     \cO[ \L_1, \m_1,\n_1,\b_1;
       R_1;\tau_1 ]}  = \delta^{[ \L_1, \m_1,\n_1,\b_1;
       R_1;\tau_1 ]}_{[ \L_2, \m_2,\n_2,\b_2; R_2;\tau_2 ]}\;
    \frac{\m_1!\n_1!\; \Dim R_1}{d_{R_1}^2} \label{eq:treediag}
\end{equation}

Now consider the one-loop correlator
\begin{equation}
   \corr{\cO^\dagger[ \L_2, \m,\n,\b_2; R_2;\tau_2 ]
     :\tr([X,Y][X^\dagger,Y^\dagger] ):\cO[ \L_1, \m,\n,\b_1;
       R_1;\tau_1 ]} \label{eq:CGbasistoget}
\end{equation}
\emph{A priori} we know that the one-loop dilatation operator will not
mix the $U(2)$ representations labelled by $\L$ and the states within
those representations labelled by $[\m,\n,\b]$ because the one-loop
dilatation operator commutes with the classical generators of $U(2)$
(and indeed of the full classical superconformal group
\cite{Beisert:2003jj})\footnote{we thank Sanjaye Ramgoolam for
  dicussions on this point}.  There is however no reason why the
$U(N)$ representations $R$ controlling the multi-trace structure
shouldn't mix and we will now analyse this using our one-loop result
\eqref{eq:finalonepoint}.

The first thing we notice, following techniques from
\cite{Brown:2007xh}, is that for a general function of a permutation
$f(\a)$
\begin{align}
  \frac{1}{n!}\sum_{\a \in S_n} B_{j \b}
  \; S^{\tau ,}{}^{ \L }_{ j }\;{}^{R}_{p }\;{}^{R}_{q}\;\;
  D_{pq}^R(\a) \sum_{\s \in S_{\m} \times S_{\n}}f(\s^{-1}\a\s) =
  \frac{\m!\n!}{n!}\sum_{\a\in S_n} B_{j \b}
  \; S^{\tau ,}{}^{ \L }_{ j }\;{}^{R}_{p }\;{}^{R}_{q}\;\;
  D_{pq}^R(\a)f(\a)
\end{align}
so that for the one-loop correlator \eqref{eq:finalonepoint} we can
absorb the $S_{\m} \times S_{\n}$ sums\footnote{another way of
  understanding this is that $\a \mapsto \s^{-1}\a\s$ for $\s \in
  S_{\m} \times S_\n$ is a symmetry of the operator $\tr(\a\; X^\m
  Y^\n )$}.

Thus if we concentrate on the $U(N)$ representation parts of equations
\eqref{eq:finalonepoint} and \eqref{eq:CGbasistoget} we find
\begin{align}
  \sum_{\a_1,\a_2 \in S_n}D_{p_1q_1}^{R_1}(\a_1)D_{p_2q_2}^{R_2}(\a_2)
  \sum_{T\,\vdash\,n+1}\chi_T(\r_1\;  \a_1 \; \r_2 \; \a_2)\Dim T 
\end{align}
If we expand the character, which is just a trace of $S_{n+1}$
representing matrices for $T$, we get
\begin{align}
  \sum_{\a_1,\a_2 \in S_n}D_{p_1q_1}^{R_1}(\a_1)D_{p_2q_2}^{R_2}(\a_2)
  \sum_{T\,\vdash\,n+1}D^T_{ab}(\r_1)D^T_{bc}( \a_1) D^T_{cd}(\r_2)D^T_{da}(\a_2)\Dim T\label{eq:reppart}
\end{align}
We can pick out the sum over $\a_1$ say
\begin{equation}
   \sum_{\a_1 \in S_n}D_{p_1q_1}^{R_1\,\vdash\, n}(\a_1)D^{T\,\vdash\, n+1}_{bc}( \a_1)
\end{equation}
$\a_1$ is in the $S_n$ subgroup of $S_{n+1}$.  As a representation of
$S_n$ the representation $T$ is reducible.  It reduces to those
$n$-box representations of $S_n$ whose Young diagrams differ by a box
from $T$.  Consider the example used in  Chapter 7 of Hamermesh
\cite{hamermesh}
\begin{equation}
\left.  T\, \tyng(6,4,4,3,2)\;\right|_{S_{18} \subset S_{19}} \to\;
T_1\, \tyng(5,4,4,3,2) \oplus T_3\, \tyng(6,4,3,3,2) \oplus T_4\,
\tyng(6,4,4,2,2) \oplus T_5\,
\tyng(6,4,4,3,1) \label{eq:Thamerexample}
\end{equation}
The index $r$ of $T_r$ labels the row from which the box was removed
from $T$.  This direct product structure is manifest for the
representation matrices constructed by Young and Yamanouchi, where the
matrix $D^T$ is block-diagonal for elements of the subgroup $\s \in
S_n \subset S_{n+1}$.  For example \eqref{eq:Thamerexample}
\begin{equation}
      D^{T\,\vdash \, n+1}(\s) = \left( 
  \begin{array}{cccc}
 D^{T_1\,\vdash \,n}(\s) & & & \\
  & D^{T_3\,\vdash \,n}(\s) & & \\
 & & D^{T_4\,\vdash \,n}(\s) & \\
 & & &  D^{T_5\,\vdash \,n}(\s)
  \end{array}
  \right) \label{eq:Thamerexplicit}
\end{equation}
For a representation $T_r$ of $S_n$ we can then apply the identity
\begin{equation}
   \sum_{\a_1 \in S_n}D_{p_1q_1}^{R_1\,\vdash \,n}(\a_1)D^{T_r\,\vdash \,n}_{bc}( \a_1) = \frac{n!}{d_{T_r}}\delta^{R_1T_r}  \delta_{p_1b} \delta_{q_1c} 
\end{equation}
This identity follows from Schur's lemma and the orthogonality of the
representing matrices.

Given the block-diagonal decomposition of $D^T$ on $\a_1$ and $\a_2$
we find that \eqref{eq:reppart} is only non-zero if $R_1 = T_r$ and
$R_2 = T_s$ for some $T$ and for some $r$ and $s$ labelling the row
from which a box is removed from $T$.  If there is no $T$ such that we
can remove a single box to reach $R_1$ and $R_2$ then the one-loop
correlator vanishes.  This is the crucial point.

If $R_1\neq R_2$ then there is at most one representation $T$ of
$S_{n+1}$ satisfying this property and we find that \eqref{eq:reppart}
becomes
\begin{align}
 \frac{n!}{d_{T_r}} \frac{n!}{d_{T_s}} D^T_{\stackrel{q_2}{s}\stackrel{p_1}{r}}(\r_1) D^T_{\stackrel{q_1}{r}\stackrel{p_2}{s}}(\r_2)\Dim T
\end{align}
The letters underneath the matrix indices indicate the sub-range of
the $d_T$ indices of $D^T$ over which the index ranges.  For example,
here $q_2$ only ranges over the $d_{T_s}$ indices of $D^T$ in the
appropriate $s$ sub-row of $D^T$ and $p_1$ only ranges over the
$d_{T_r}$ indices in the $r$ sub-column (see for example the matrix in
\eqref{eq:Thamerexplicit})\footnote{if we want to be more fancy $s$ is
  the first number in the Yamanouchi symbol for the index of $T$ and
  $q_2$ is the rest of the symbol for $T_s$}.  Thus for
$D^T_{\stackrel{q_2}{s}\stackrel{p_1}{r}}(\r_1)$ $q_2$ and $p_1$ label
elements in an off-diagonal sub-block of $D^T$.  This does not vanish
because $\rho_1$ is a generic element of $S_{n+1}$ not in its $S_n$
subgroup.

So if there exists a $T$ for which $R_1 = T_r$ and $R_2 = T_s$ and
$R_1 \neq R_2$
\begin{align}
&   \corr{\cO^\dagger[ \L_2, \m,\n,\b_2; T_s;\tau_2 ]
     :\tr([X,Y][X^\dagger,Y^\dagger] ):\cO[ \L_1, \m,\n,\b_1;
       T_r;\tau_1 ]}  \nn \\
& =  \frac{\m\n\m!\n!}{d_{T_r}d_{T_s}} B_{j_1 \b_1}
  \; S^{\tau_1,}{}^{ \L_1 }_{ j_1}\;{}^{T_r}_{p_1}\;{}^{T_r}_{q_1}\;  B_{j_2 \b_2}
  \; S^{\tau_2,}{}^{ \L_2 }_{ j_2}\;{}^{T_s}_{p_2}\;{}^{T_s}_{q_2}\;\sum_{\r_1,\r_2\in S_{n+1}} h(\r_1,\r_2) D^T_{\stackrel{q_2}{s}\stackrel{p_1}{r}}(\r_1) D^T_{\stackrel{q_1}{r}\stackrel{p_2}{s}}(\r_2)\Dim T
\end{align}
If we use the more symmetric expression for $h$ in \eqref{eq:fullh2}
then we can use identity \eqref{Hamer186} from Appendix Section
\ref{sec:convandform} to get
\begin{align}
& -\frac{\m\n\m!\n!}{d_{T_r}d_{T_s}} B_{j_1 \b_1}
  \; S^{\tau_1,}{}^{ \L_1 }_{ k_1}\;{}^{T_r}_{p_1}\;{}^{T_r}_{q_1}\;  B_{j_2 \b_2}
  \; S^{\tau_2,}{}^{ \L_2 }_{
    k_2}\;{}^{T_s}_{p_2}\;{}^{T_s}_{q_2}\;\nn\\
& D^{\L_1}_{j_1k_1}(1-(\m,n))  \;D^{\L_2}_{j_2k_2}(1-(\m,n)) \;
  D^T_{\stackrel{q_2}{s}\stackrel{p_1}{r}}((\m,n+1))
  D^T_{\stackrel{q_1}{r}\stackrel{p_2}{s}}((n,n+1))\Dim
  T \label{eq:mixingmatrix}
\end{align}
This expression nicely encodes the vanishing of the one-loop
correlator for the half-BPS operators transforming in the symmetric
representation of the flavour group (for $\L =
\tyoung(\ \ \scdots\ )\,$, $D^{\L}(\s) = 1 \;\;\forall \s$).

Some hints on how to simplify this expression further, and how one
might extract explicitly the orthogonality of $U(2)$ representations,
is given in Appendix Section \ref{sec:further}.

If $R_1 = R_2\equiv R$ then we must sum over all the representations
$T$ of $S_{n+1}$ with $T_r = R$
\begin{align}
&   \corr{\cO^\dagger[ \L_2, \m,\n,\b_2; R;\tau_2 ]
     :\tr([X,Y][X^\dagger,Y^\dagger] ):\cO[ \L_1, \m,\n,\b_1;
       R;\tau_1 ]}  \nn \\
& = \sum_{T\textrm{ s.t. } R =T_r} \frac{\m\n\m!\n!}{d_{T_r}^2} B_{j_1 \b_1}
  \; S^{\tau_1,}{}^{ \L_1 }_{ j_1}\;{}^{T_r}_{p_1}\;{}^{T_r}_{q_1}\;  B_{j_2 \b_2}
  \; S^{\tau_2,}{}^{ \L_2 }_{
    j_2}\;{}^{T_r}_{p_2}\;{}^{T_r}_{q_2}\;\sum_{\r_1,\r_2\in S_{n+1}}
  h(\r_1,\r_2) D^T_{\stackrel{q_2}{r}\stackrel{p_1}{r}}(\r_1)
  D^T_{\stackrel{q_1}{r}\stackrel{p_2}{r}}(\r_2)\Dim T \nn
\end{align}

An example of these mixing properties is worked out for $\L =
\tyng(2,2)$ in Appendix Section \ref{sec:example}.

Some general comments:
\begin{itemize}
\item We can interpret the $U(N)$ representation $T\,\vdash \,n+1$ as
  an intermediate channel through which the operators mix via the
  `overlapping' of $R_1\,\vdash \,n$ and $R_2\,\vdash \,n$ with $T$.
\item Given that smaller Young diagrams are more likely to be related
  to each other by moving a box than larger diagrams, mixing at one
  loop is much more likely for smaller representations than larger
  ones.  Larger ones can be considered practically diagonal at 1-loop
  (but not at higher loops, see Section \ref{sec:extensions}).
\end{itemize}

\subsection{the dilatation operator}

We can now apply this analysis to the one-loop dilatation operator.
\begin{equation}
  \Delta^{(1)}\; \cO[\L,\m,\n,\b; R; \tau] = \sum_{S,\tau'}\;C^{R,\tau}_{S,\tau'}\; \cO[\L,\m,\n,\b; S; \tau']
\end{equation}
$S$ must be obtainable by removing a box from $R$ and then putting it
back somewhere.  We can obtain the matrix $C^{R,\tau}_{S,\tau'}$ by
reverse-engineering the one-loop mixing \eqref{eq:mixingmatrix} using
the tree-diagonality of the Clebsch-Gordan basis \eqref{eq:treediag}.
We can see for example that for $R \neq S$ which mix via $T \vdash
n+1$ we can factor out the $N$ dependence
\begin{align}
  C^{R;\tau}_{S;\tau'} = &
 -\m\n\frac{ d_{S}}{d_{R}} \frac{\Dim
  T}{\Dim S} B_{j_1 \b}
  \; S^{\tau,}{}^{ \L}_{ k_1}\;{}^{R}_{p_1}\;{}^{R}_{q_1}\;  B_{j_2 \b}
  \; S^{\tau',}{}^{ \L }_{
    k_2}\;{}^{S}_{p_2}\;{}^{S}_{q_2}\;\nn\\
& D^{\L}_{j_1k_1}(1-(\m,n))  \;D^{\L}_{j_2k_2}(1-(\m,n)) \;
  D^T_{\stackrel{q_2}{s}\stackrel{p_1}{r}}((\m,n+1))
  D^T_{\stackrel{q_1}{r}\stackrel{p_2}{s}}((n,n+1)) \nn \\
\propto &\;\;\frac{\Dim T}{\Dim S} \;\;\propto \;\;N-i + j
\end{align}
where $i$ labels the row coordinate and $j$ the column coordinate of
the box $R$ has that $S$ doesn't (see equation \eqref{eq:dimR}).

The kernel of this map provides the $\frac{1}{4}$-BPS operators
\cite{Ryzhov:2001bp}\cite{D'Hoker:2003vf}, but we have no further
insight on how to obtain a pleasing group theoretic expression for
these operators beyond the hints given in \cite{Brown:2007xh}
concerning the dual basis \cite{Brown:2007bb}\cite{Brown:2006zk}.
Something like the dual basis seems particularly relevant given that
it arose in the $SU(N)$ context
\cite{deMelloKoch:2004ws}\cite{Brown:2007bb} from knocking boxes off
representations to differentiate Schur polynomials.

\section{higher loops and other sectors}\label{sec:extensions}

If we assume that higher $\ell$-loop contributions to the correlator
can always be written in terms of an effective vertex like
\eqref{eq:normalorder} (it works for two loops \cite{Beisert:2003tq})
then we guess that they can be written in terms of $S_{n + \ell}$ and
$U(N)$ group theory
\begin{equation}
\sum_{\s,\tau \in S_\m \times S_\n}\sum_{\r_1,\r_2\in S_{n+\ell}}
h_{\ell}(\r_1,\r_2) \sum_{T\,\vdash \,n+\ell}\chi_T(\r_1\; \s^{-1} \a_1 \s \; \r_2
\;\tau^{-1} \a_2 \tau )\Dim T
\end{equation}
$h_\ell(\r_1,\r_2)$ only takes non-zero values on a few permutations
of $\ell +1$ of the $\{1, \dots n\}$ indices (where the derivative
acts) and the $n+1$, \dots $n + \ell$ indices.  The $\sigma$ and $\tau$
construction permutes the $X$'s and $Y$'s for the product rule.

This guess is informed by the leading planar $N^{n+\ell}$ contribution
to the $\ell$-loop term, which is provided by the large $N$ behaviour
of $\Dim T$ when $T$ has $n+\ell$ boxes (see equation
\eqref{eq:dimR}).

As a consequence of this structure $\cO[ \L_1, \m,\n,\b_1;R_1;\tau_1
]$ and $\cO[ \L_2, \m,\n,\b_2; R_2;\tau_2 ]$ can only mix at $\ell$
loops if we can reach the same $(n+\ell)$-box Young diagram $T$ by
adding $\ell$ boxes to each of the $U(N)$ representations $R_1$ and
$R_2$.

An alternative way of saying this is that if two $U(N)$
representations $R_1$ and $R_2$ have $k$ boxes in the same position
then they can first mix at $n-k$ loops, since we have enough boxes to
add to $R_1$ to reproduce the shape of $R_2$.

This means that all operators of length $n$ can mix at $n-1$ loops,
because all diagrams share the first box in the upper lefthand corner.

We have focused here on the $U(2) \subset SU(4) \subset PSU(2,2|4)$
sector of the full symmetry group of $\cN = 4$.  It seems fairly
obvious that this work extends to $U(3)$ because the effective vertex
gains similar terms to the $U(2)$ vertex and the basis of
\cite{Brown:2007xh} accommodates a general $U(M)$ flavour symmetry;
the remaining sectors \cite{Beisert:2003jj} would require more work,
especially given that the basis constructed in \cite{Brown:2007xh}
doesn't extend there yet.  It would be particularly interesting to
extend the work of \cite{Janik:2007pm} and understand the counting of
sixteenth-BPS operators at one loop in the non-planar limit, and hence
gain an understanding of black hole entropy via AdS/CFT.

There are satisfying group-theoretic expressions for extremal
higher-point correlators of the Clebsch-Gordan operators at tree level
\cite{Brown:2007xh}.  It would be interesting to see how much of this
structure survives at one loop.

Finally we point out that another complete basis in the $U(2)$ sector, the restricted Schur
polynomials, have neat tree-level two-point functions and their
one-loop properties have been studied \cite{de Mello Koch:2007uu}\cite{de Mello Koch:2007uv}\cite{Bekker:2007ea}\cite{Bhattacharyya:2008rb}.

\section{discussion}

The main motivation for studying these operators and their mixing is
that $\cN=4$ super Yang-Mills has a dual string theory on an $AdS_5
\times S^5$ background
\cite{Maldacena:1997re}\cite{Gubser:1998bc}\cite{Witten:1998qj}.  We
give here some techniques that allow us better control of the regime
where the length of operators is arbitrary, $\l$ is non-trivial and
$N$ is finite, the regime where the `strong' Maldacena conjecture
might hold beyond the planar 't Hooft limit.

We have no clear idea what the tree-diagonal operators constructed in
\cite{Brown:2007xh} correspond to on the string theory side.  They are
not eigenstates of the one-loop dilatation operator, but their limited
mixing might pave the way for such a diagonalisation.  The BPS
operators map to giant graviton branes when the operators are large
\cite{Mikhailov:2000ya}\cite{Beasley:2002xv}\cite{Biswas:2006tj}\cite{Mandal:2006tk}.
Some hints on how to obtain these operators from the Clebsch-Gordan
basis were given in \cite{Brown:2007xh}.

On the string side splitting of strings is suppressed by $g_s \sim
1/N$.  One lesson perhaps is that it is fruitful to think in terms of
Young diagrams gaining and losing boxes as well as in terms of traces
splitting and joining.  An advantage of the Young diagram methods is
that the finite $N$ constraint is clear in terms of a limit on the
number of rows.  It would be interesting to understand how this
constraint \cite{Maldacena:1998bw} is implemented for general string
states, particularly given that it is reminiscent of the level cutoff
of Wess-Zumino-Witten models \cite{Witten:1983ar}.

Representation theory and Schur-Weyl duality played an important part
in our understanding of 2d Yang-Mills and its string dual
\cite{Gross:1992tu}\cite{Gross:1993hu}\cite{Cordes:1994fc}.  We hope
that Schur-Weyl duality, and the interplay between the gauge group and
the symmetry group, will provide vital clues for our understanding of
$d=4, \cN=4$ supersymmetric Yang-Mills and the string on $AdS_5 \times
S^5$.

\vspace{2cm}

\noindent{\bf Acknowledgements}\; We thank Paul Heslop, Robert de
Mello Koch, Sanjaye Ramgoolam, Rodolfo Russo and Konstantinos Zoubos
for valuable discussions.  We also thank Se\'an Murray for help with
the grisly, noisome typesetting.  TWB is on an STFC studentship.

\vspace{5mm}

\begin{appendix}

\section{conventions and formulae}\label{sec:convandform}

$R\vdash n$ is an irreducible representation of $S_n$ and also of
$U(N)$.  It can be drawn as a Young diagram with $n$ boxes;
representations of $U(N)$ have at most $N$ rows.

$d_R = \frac{n!}{\prod_{i,j}h_{i,j}}$ is the dimension of the
symmetric group representation $R$, where $h_{i,j}$ is the hook length
for the box in the $i$th row and $j$th column.

$\Dim R$ is the dimension of the unitary group $U(N)$ representation
$R$, given by
\begin{equation}
  \Dim R = \prod_{(i,j)\in R} \frac{N-i+j}{h_{i,j}} \label{eq:dimR}
\end{equation}
Again $i$ labels the row coordinate and $j$ the column coordinate of
each box in $R$.

The $S_n$ Clebsch-Gordan coefficients satisfy for a permutation $\s
\in S_n$
\begin{align}
 \sum_{j,l}  D^S_{ij} (\s) D^T_{kl} (\s)\;S^{ \tau_R,}{}^{ R}_{s}\;{}^S_j\;{}^T_l & =
  \sum_t D^R_{ts} (\s)\;
  S^{ \tau_R,}{}^{ R}_{ t}\;{}^S_i\;{}^T_k \label{Hamer186}
\end{align}
This tells us how to obtain matrix elements from the symmetric group
inner product $R \in S \otimes T$.  $\tau_R$ labels the multiplicity
of $R$ in $S \otimes T$.

\section{diagrammatics}\label{sec:diagrammatics}

\begin{figure}[h]
\begin{center}
\includegraphics{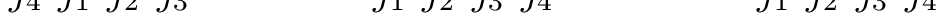}
\caption{from delta functions to diagrams to
  permutations} \label{fig:diagrammatics}
\end{center}
\end{figure}

\begin{figure}[h]
\begin{center}
\includegraphics{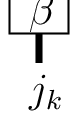}
\caption{permutations in series; thick lines represent many strands} \label{fig:thickstrand}
\end{center}
\end{figure}

Diagrammatics \cite{Corley:2002mj} encode the 't Hooft double-line
indices.  We follow the index lines with delta functions and
permutations, see for example Figure \ref{fig:diagrammatics}.
We read the permutations in the diagrams from the top down.  This is
also illustrated in Figure \ref{fig:thickstrand}, where we remember
that in the permutation $\b\a$ we read from right to left, so that
$\a$ acts first followed by $\b$.
Also in Figure \ref{fig:thickstrand} we clump several strands labelled
by $k$ into a single thick strand, for clarity.

\begin{figure}[h]
\begin{center}
\includegraphics{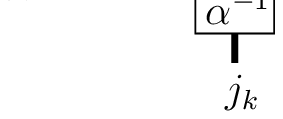}
\caption{permutations on the upper index} \label{fig:upper}
\end{center}
\end{figure}

If we write down a series of delta functions we can always alter the
order in which we write them down with any $\s \in S_n$, given that
they are just numbers
\begin{equation}
  \delta^{i_{1}}_{j_{\a(1)}} \cdots   \delta^{i_{n}}_{j_{\a(n)}} =
          \delta^{i_{\s(1)}}_{j_{\a\s(1)}} \cdots
              \delta^{i_{\s(n)}}_{j_{\a\s(n)}}
\end{equation}
This allows us to deal with permutations on the upper index, see
Figure \ref{fig:upper}.

If we have $\delta^{i_{\a(k)}}_{j_{\b(k)}}$ and we set $j_k =
i_{\s(k)}$ then we get 
\begin{equation}
\delta^{i_{\a(k)}}_{j_{\b(k)}} \delta^{j_{k}}_{i_{\s(k)}} =
\delta^{i_{\a\b^{-1}(k)}}_{j_{k}} \delta^{j_{k}}_{i_{\s(k)}} =
\delta^{i_{\a\b^{-1}(k)}}_{i_{\s(k)}} = \delta^{i_{\a(k)}}_{i_{\s\b(k)}} 
\end{equation}

\section{symmetric group representation matrices}\label{sec:symmetricgroup}

Here we briefly review the Young-Yamanouchi construction of real
orthogonal representing matrices for an $S_n$ representation $T$
\cite{yamanouchi}, which is summarised in Hamermesh \cite{hamermesh}.

The matrices are constructed recursively: we assume that we know all
the representation matrices for all the representations of $S_k$ for
$k < n$.  We also know that on elements of the subgroup $S_{n-1}
\subset S_n$ the representation $T$ reduces to a sum of those
irreducible representations of $S_{n-1}$ that have one box removed
from $T$ (see for example equations \eqref{eq:Thamerexample} and
\eqref{eq:Thamerexplicit}).  Given that we know all the representation
matrices for all of $S_{n-1}$ we know the form of the representation
matrices for $T$ on $S_{n-1} \subset S_n$.

To reach those permutations that also act on the last object, all we
need to know in addition is the matrix for $(n-1,n)$, $D^T((n-1,n))$.
To obtain this, we observe that this matrix commutes with all the
matrices for the subgroup $S_{n-2} \subset S_n$, since they are
permuting separate groups of objects.  We can then use Schur's lemmas
to obtain $D^T((n-1,n))$.

\begin{align}
 \textrm{Type I:}& \quad\quad T_{11}\;\;
 \tyoung(\ \ \ \ \times\times,\ \ \ \ ,\ \ \ \ ,\ \ \ ,\ \ ) \quad
 \textrm{and} \quad  T_{55}\;\;
 \tyoung(\ \ \ \ \ \ ,\ \ \ \ ,\ \ \ \ ,\ \ \ ,\times\times) \nn\\
 \textrm{Type II:}& \quad\quad T_{13} = T_{31}\;\;
 \tyoung(\ \ \ \ \ \times,\ \ \ \ ,\ \ \ \times,\ \ \ ,\ \ )\;,  \quad
  T_{34} = T_{43}\;\;
 \tyoung(\ \ \ \ \ \ ,\ \ \ \ ,\ \ \ \times,\ \ \times,\ \ )\;, \quad \cdots \nn\\
 \textrm{Type III:}& \quad\quad T_{32}\;\;
 \tyoung(\ \ \ \ \ \ ,\ \ \ \times,\ \ \ \times,\ \ \ ,\ \ )
\end{align}

To get the representing matrices of $T$ on $S_{n-2} \subset S_n$, we
must reduce $T$ by knocking off two boxes.  We label these irreps of
$S_{n-2}$ by $T_{rs}$ where $r$ is the row from which the first box is
knocked, $s$ the second.  There are three different situations when we
knock off two boxes, called Type I, II and III.  These are exhibited
for the example given in equation \eqref{eq:Thamerexample}.

For Type I and Type III the second box can only be knocked off
\emph{after} the first one: Type I is when the second box is to the
left of the first on the same row; Type III is when the second box is
above the first on the same column.  For Type II both boxes can be
knocked off independently and $T_{rs} = T_{sr}$.

This reduction of $S_n$ representations on subgroups is also called
\emph{branching}.

\section{further analysis of the matrices}\label{sec:further}

Here we analyse in more detail the one-loop mixing of the
Clebsch-Gordan basis for $R_1 = T_r$ and $R_2 = T_s$ and $r \neq s$
given in \eqref{eq:mixingmatrix}.

\begin{figure}[h]
\begin{center}
\begin{picture}(150,100)
\put(0,50){\tyng(3,2)}
\put(35,54){\vector(2,1){20}}
\put(35,46){\vector(2,-1){20}}
\put(35,60){\vector(2,3){22}}
\put(35,40){\vector(2,-3){22}}
\put(60,65){\tyng(2,2)}
\put(60,35){\tyng(3,1)}
\put(60,95){$\cdots$}
\put(60,0){$\cdots$}
\put(120,50){\tyng(2,1)}
\put(120,90){$\cdots$}
\put(120,5){$\cdots$}
\put(90,63){\vector(2,-1){20}}
\put(90,37){\vector(2,1){20}}
\put(90,67){\vector(1,1){22}}
\put(90,33){\vector(1,-1){22}}
\end{picture}
\end{center}
\caption{restriction pattern for $S_{n+1}\to S_{n} \to S_{n-1}$}  \label{fig:tonminus2}
\end{figure}

It turns out, given the recursive construction of the representing
matrices (see Appendix Section \ref{sec:symmetricgroup}), that we know
$D^T_{\stackrel{q_1}{r}\stackrel{p_2}{s}}((n,n+1))$ exactly.  If we
further restrict $T$ to $S_{n-1}$ then the representation reduces to
Young diagrams with two boxes removed from $T$.  $T_{rs} = T_{sr}$ is
the common $S_{n-1}$ Young diagram obtained when boxes are removed
both from the $r$th and $s$th rows (see Figure \ref{fig:tonminus2}).
It is Type II because the boxes can be removed independently.  Because
$(n,n+1)$ commutes with all elements of $S_{n-1}$,
$D^T_{\stackrel{q_1}{r}\stackrel{p_2}{s}}((n,n+1))$ is only non-zero
in the case
\begin{equation}
  D^T_{\stackrel{q_1}{rs}\,\stackrel{p_2}{sr}}((n,n+1)) =
  \frac{\sqrt{\tau_{rs,rs}^2 - 1}}{|\tau_{rs,rs}|} E_{rs,sr}
\end{equation}
where $E_{rs,sr}$ is the identity matrix.  If the row lengths of $T$
are given by $t_r$ then $\tau_{rs,rs}$ is\footnote{$\tau_{rs,rs}$ is
  also known as the \emph{axial distance}}
\begin{equation}
  \tau_{rs,rs} =( t_r -r) - (t_s - s)
\end{equation}

Unfortunately we can't work the same magic on
$D^T_{\stackrel{q_2}{s}\stackrel{p_1}{r}}((\m,n+1))$.

There are also branching-type recursive relations for the
Clebsch-Gordan coefficients (see the end of Chapter 7 of Hamermesh
\cite{hamermesh}).

Given that we know \eqref{eq:mixingmatrix} is diagonal in the $U(2)$
states, this may imply non-trivial identities for these symmetric
group reduction formulae.

\section{example}\label{sec:example}

We consider the case with $U(2)$ representation $\L = \tyng(2,2)$ and
field content $XXYY$.  This must be a highest weight state of $\L$
because the field content matches the rows of $\L$.  Thus $\b$ is
unique.

The three allowed $U(N)$ representations are $R =
\tyng(2,2),\tyng(3,1), \tyng(2,1,1)$, for which $\L$ only appears once
in the symmetric group inner product $R \otimes R$.

Here $\Phi_r \Phi^r = \e_{rs} \Phi^r \Phi^s = [X,Y]$.
\begin{align}
    \cO\left[\L = \tyng(2,2); R = \tyng(2,2)  \right]
 =  \frac{1}{12\sqrt{2}} \left[\tr(\Phi_r \Phi_s)\tr(\Phi^r)\tr(\Phi^s) +
 \tr(\Phi_r\Phi^r\Phi_s \Phi^s) \right]
\end{align}

\begin{align}
     \cO\left[\L = \tyng(2,2); R = \tyng(3,1)  \right]
 =  \frac{1}{12\sqrt{6}} \left[ \tr(\Phi_r \Phi_s)
 \tr(\Phi^r)\tr(\Phi^s) + \tr(\Phi_r \Phi_s)\tr(\Phi^r \Phi^s) -
 \tr(\Phi_r\Phi^r\Phi_s \Phi^s) \right] 
\end{align}

\begin{align}
   \cO\left[\L = \tyng(2,2); R = \tyng(2,1,1)  \right]
 =  \frac{1}{12\sqrt{6 }} \left[ \tr(\Phi_r \Phi_s)
 \tr(\Phi^r)\tr(\Phi^s) - \tr(\Phi_r \Phi_s)\tr(\Phi^r \Phi^s) -
 \tr(\Phi_r\Phi^r\Phi_s \Phi^s) \right] 
\end{align}

The tree level correlator is diagonal
\begin{align}
&  \left(
  \begin{array}{ccc}
    \frac{1}{12}N^2(N^2-1) & & \\
     &  \frac{1}{18}N(N^2-1)(N+2) & \\
    & &  \frac{1}{18}N(N^2-1)(N-2)
  \end{array} \right) \nn\\
&\nn \\
&  =   \left(
  \begin{array}{ccc}
\Dim\, \tyng(2,2) &  & \\
&&\\
     &  \frac{4}{9} \Dim\, \tyng(3,1)& \\
&&\\
     & & \frac{4}{9} \Dim\, \tyng(2,1,1)
  \end{array} \right)
\end{align}

At one loop everything mixes
\begin{align}
&  \left(
  \begin{array}{ccc}
    \frac{1}{4}N^3(1-N^2) & \frac{1}{4\sqrt{3}}N^2(N^2-1)(N+2) &\frac{1}{4\sqrt{3}}N^2(N^2-1)(N-2) \\
    \frac{1}{4\sqrt{3}}N^2(N^2-1)(N+2)  &  \frac{1}{12}N(1-N^2)(N+2)^2 & \frac{1}{12}N(1-N^2)(N^2-4) \\
   \frac{1}{4\sqrt{3}}N^2(N^2-1)(N-2) & \frac{1}{12}N(1-N^2)(N^2-4) & \frac{1}{12}N(1-N^2)(N-2)^2
  \end{array} \right) \nn\\
&\nn \\
&  =   \left(
  \begin{array}{ccc}
    -3N\Dim \, \tyng(2,2) & 2\sqrt{3}\Dim\, \tyng(3,2) &   2\sqrt{3}\Dim\,
    \tyng(2,2,1) \\
 &&\\
 2\sqrt{3}\Dim\, \tyng(3,2)    &  -\frac{2}{3}(N+2)\Dim\, \tyng(3,1) & -\frac{5}{3} \Dim\, \tyng(3,1,1) \\
&&\\
  2\sqrt{3}\Dim\,
    \tyng(2,2,1)  & -\frac{5}{3} \Dim\, \tyng(3,1,1) &
    -\frac{2}{3}(N-2) \Dim\, \tyng(2,1,1)
  \end{array} \right)
\end{align}
The diagonal terms seem to be the dimension of the irrep. enhanced by
the contribution for a specific box, furthest from the top left.

\section{code}\label{sec:code}

All correlators at tree level and one loop can be checked with the
correlator program written in \href{http://www.python.org/}{python}
and released under the \href{http://www.gnu.org/copyleft/gpl.html}{GNU
  General Public Licence} at \url{http://www.nworbmot.org/code/}.

\end{appendix}


\begin{thebibliography}{99}


\bibitem{Brown:2007xh}
  T.~W.~Brown, P.~J.~Heslop and S.~Ramgoolam,
  ``Diagonal multi-matrix correlators and BPS operators in N=4 SYM,''
  arXiv:0711.0176 [hep-th].

\bibitem{'t Hooft:1973jz}
  G.~'t Hooft,
  ``A planar diagram theory for strong interactions,''
  Nucl.\ Phys.\  B {\bf 72} (1974) 461.

\bibitem{Corley:2001zk}
  S.~Corley, A.~Jevicki and S.~Ramgoolam,
  ``Exact correlators of giant gravitons from dual N = 4 SYM theory,''
  Adv.\ Theor.\ Math.\ Phys.\  {\bf 5} (2002) 809
  [arXiv:hep-th/0111222].

\bibitem{Corley:2002mj}
  S.~Corley and S.~Ramgoolam,
  ``Finite factorization equations and sum rules for BPS correlators in  N = 4
  SYM theory,''
  Nucl.\ Phys.\  B {\bf 641} (2002) 131
  [arXiv:hep-th/0205221].

\bibitem{Constable:2002hw}
  N.~R.~Constable, D.~Z.~Freedman, M.~Headrick, S.~Minwalla, L.~Motl, A.~Postnikov and W.~Skiba,
  ``PP-wave string interactions from perturbative Yang-Mills theory,''
  JHEP {\bf 0207} (2002) 017
  [arXiv:hep-th/0205089].

\bibitem{Beisert:2002bb}
  N.~Beisert, C.~Kristjansen, J.~Plefka, G.~W.~Semenoff and M.~Staudacher,
  ``BMN correlators and operator mixing in N = 4 super Yang-Mills theory,''
  Nucl.\ Phys.\  B {\bf 650} (2003) 125
  [arXiv:hep-th/0208178].

\bibitem{Gross:2002mh}
  D.~J.~Gross, A.~Mikhailov and R.~Roiban,
  ``A calculation of the plane wave string Hamiltonian from N = 4
  super-Yang-Mills theory,''
  JHEP {\bf 0305} (2003) 025
  [arXiv:hep-th/0208231].

\bibitem{Janik:2002bd}
  R.~A.~Janik,
  ``BMN operators and string field theory,''
  Phys.\ Lett.\  B {\bf 549} (2002) 237
  [arXiv:hep-th/0209263].

\bibitem{Beisert:2002ff}
  N.~Beisert, C.~Kristjansen, J.~Plefka and M.~Staudacher,
  ``BMN gauge theory as a quantum mechanical system,''
  Phys.\ Lett.\  B {\bf 558} (2003) 229
  [arXiv:hep-th/0212269].

\bibitem{Minahan:2002ve}
  J.~A.~Minahan and K.~Zarembo,
  ``The Bethe-ansatz for N = 4 super Yang-Mills,''
  JHEP {\bf 0303} (2003) 013
  [arXiv:hep-th/0212208].

\bibitem{Beisert:2003tq}
  N.~Beisert, C.~Kristjansen and M.~Staudacher,
  ``The dilatation operator of N = 4 super Yang-Mills theory,''
  Nucl.\ Phys.\  B {\bf 664} (2003) 131
  [arXiv:hep-th/0303060].

\bibitem{Beisert:2003yb}
  N.~Beisert and M.~Staudacher,
  ``The N = 4 SYM integrable super spin chain,''
  Nucl.\ Phys.\  B {\bf 670} (2003) 439
  [arXiv:hep-th/0307042].

\bibitem{Bellucci:2004ru}
  S.~Bellucci, P.~Y.~Casteill, J.~F.~Morales and C.~Sochichiu,
  ``Spin bit models from non-planar N = 4 SYM,''
  Nucl.\ Phys.\  B {\bf 699} (2004) 151
  [arXiv:hep-th/0404066].

\bibitem{Arutyunov:2000ima}
  G.~Arutyunov and S.~Frolov,
  ``On the correspondence between gravity fields and CFT operators,''
  JHEP {\bf 0004} (2000) 017
  [arXiv:hep-th/0003038].

\bibitem{Arutyunov:2002rs}
  G.~Arutyunov, S.~Penati, A.~C.~Petkou, A.~Santambrogio and E.~Sokatchev,
  ``Non-protected operators in N = 4 SYM and multiparticle states of AdS(5)
  SUGRA,''
  Nucl.\ Phys.\  B {\bf 643} (2002) 49
  [arXiv:hep-th/0206020].

\bibitem{Bianchi:2002rw}
  M.~Bianchi, B.~Eden, G.~Rossi and Y.~S.~Stanev,
  ``On operator mixing in N = 4 SYM,''
  Nucl.\ Phys.\  B {\bf 646} (2002) 69
  [arXiv:hep-th/0205321].

\bibitem{Constable:2002vq}
  N.~R.~Constable, D.~Z.~Freedman, M.~Headrick and S.~Minwalla,
  ``Operator mixing and the BMN correspondence,''
  JHEP {\bf 0210} (2002) 068
  [arXiv:hep-th/0209002].

\bibitem{hamermesh}
  M.~Hamermesh,
  ``Group Theory and its Application to Physical Problems,''
  Addison-Wesley Publishing Company (1962).

\bibitem{yamanouchi}
  T.~Yamanouchi,
  Proc.\ Phys.\ Math.\ Soc.\ Japan {\bf 19} (1937) 436.

\bibitem{Beisert:2003jj}
  N.~Beisert,
  ``The complete one-loop dilatation operator of N = 4 super Yang-Mills
  theory,''
  Nucl.\ Phys.\  B {\bf 676} (2004) 3
  [arXiv:hep-th/0307015].

\bibitem{Ryzhov:2001bp}
  A.~V.~Ryzhov,
  ``Quarter BPS operators in N = 4 SYM,''
  JHEP {\bf 0111} (2001) 046
  [arXiv:hep-th/0109064].

\bibitem{D'Hoker:2003vf}
  E.~D'Hoker, P.~Heslop, P.~Howe and A.~V.~Ryzhov,
  ``Systematics of quarter BPS operators in N = 4 SYM,''
  JHEP {\bf 0304} (2003) 038
  [arXiv:hep-th/0301104].

\bibitem{Brown:2007bb}
  T.~W.~Brown,
  ``Half-BPS SU(N) correlators in N = 4 SYM,''
  [arXiv:hep-th/0703202].

\bibitem{Brown:2006zk}
  T.~Brown, R.~de Mello Koch, S.~Ramgoolam and N.~Toumbas,
  ``Correlators, probabilities and topologies in N = 4 SYM,''
  JHEP {\bf 0703} (2007) 072
  [arXiv:hep-th/0611290].

\bibitem{deMelloKoch:2004ws}
  R.~de Mello Koch and R.~Gwyn,
  ``Giant graviton correlators from dual SU(N) super Yang-Mills theory,''
  JHEP {\bf 0411} (2004) 081
  [arXiv:hep-th/0410236].

\bibitem{Janik:2007pm}
  R.~A.~Janik and M.~Trzetrzelewski,
  ``Supergravitons from one loop perturbative N=4 SYM,''
  arXiv:0712.2714 [hep-th].


\bibitem{de Mello Koch:2007uu}
  R.~de Mello Koch, J.~Smolic and M.~Smolic,
  ``Giant Gravitons - with Strings Attached (I),''
  JHEP {\bf 0706} (2007) 074
  [arXiv:hep-th/0701066].

\bibitem{de Mello Koch:2007uv}
  R.~de Mello Koch, J.~Smolic and M.~Smolic,
  ``Giant Gravitons - with Strings Attached (II),''
  JHEP {\bf 0709} (2007) 049
  [arXiv:hep-th/0701067].

\bibitem{Bekker:2007ea}
  D.~Bekker, R.~de Mello Koch and M.~Stephanou,
  ``Giant Gravitons - with Strings Attached (III),''
  JHEP {\bf 0802} (2008) 029
  arXiv:0710.5372 [hep-th].

\bibitem{Bhattacharyya:2008rb}
  R.~Bhattacharyya, S.~Collins and R.~d.~M.~Koch,
  ``Exact Multi-Matrix Correlators,''
  JHEP {\bf 0803} (2008) 044
  arXiv:0801.2061 [hep-th].

\bibitem{Maldacena:1997re}
  J.~M.~Maldacena,
  ``The large N limit of superconformal field theories and supergravity,''
  Adv.\ Theor.\ Math.\ Phys.\  {\bf 2} (1998) 231
  [Int.\ J.\ Theor.\ Phys.\  {\bf 38} (1999) 1113]
  [arXiv:hep-th/9711200].

\bibitem{Gubser:1998bc}
  S.~S.~Gubser, I.~R.~Klebanov and A.~M.~Polyakov,
  ``Gauge theory correlators from non-critical string theory,''
  Phys.\ Lett.\  B {\bf 428} (1998) 105
  [arXiv:hep-th/9802109].

\bibitem{Witten:1998qj}
  E.~Witten,
  ``Anti-de Sitter space and holography,''
  Adv.\ Theor.\ Math.\ Phys.\  {\bf 2} (1998) 253
  [arXiv:hep-th/9802150].

\bibitem{Mikhailov:2000ya}
  A.~Mikhailov,
  ``Giant gravitons from holomorphic surfaces,''
  JHEP {\bf 0011} (2000) 027
  [arXiv:hep-th/0010206].

\bibitem{Beasley:2002xv}
  C.~E.~Beasley,
  ``BPS branes from baryons,''
  JHEP {\bf 0211} (2002) 015
  [arXiv:hep-th/0207125].

\bibitem{Biswas:2006tj}
  I.~Biswas, D.~Gaiotto, S.~Lahiri and S.~Minwalla,
  ``Supersymmetric states of N = 4 Yang-Mills from giant gravitons,''
  [arXiv:hep-th/0606087].

\bibitem{Mandal:2006tk}
  G.~Mandal and N.~V.~Suryanarayana,
  ``Counting 1/8-BPS dual-giants,''
  JHEP {\bf 0703} (2007) 031
  [arXiv:hep-th/0606088].

\bibitem{Maldacena:1998bw}
  J.~M.~Maldacena and A.~Strominger,
  ``AdS(3) black holes and a stringy exclusion principle,''
  JHEP {\bf 9812} (1998) 005
  [arXiv:hep-th/9804085].

\bibitem{Witten:1983ar}
  E.~Witten,
  ``Nonabelian bosonization in two dimensions,''
  Commun.\ Math.\ Phys.\  {\bf 92} (1984) 455.

\bibitem{Gross:1992tu}
  D.~J.~Gross,
  ``Two-dimensional QCD as a string theory,''
  Nucl.\ Phys.\  B {\bf 400} (1993) 161
  [arXiv:hep-th/9212149].

\bibitem{Gross:1993hu}
  D.~J.~Gross and W.~Taylor,
  ``Two-dimensional QCD is a string theory,''
  Nucl.\ Phys.\  B {\bf 400} (1993) 181
  [arXiv:hep-th/9301068].

\bibitem{Cordes:1994fc}
  S.~Cordes, G.~W.~Moore and S.~Ramgoolam,
  ``Lectures On 2-D Yang-Mills Theory, Equivariant Cohomology And Topological
  Field Theories,''
  Nucl.\ Phys.\ Proc.\ Suppl.\  {\bf 41} (1995) 184
  [arXiv:hep-th/9411210].

\end{thebibliography}
\end{document}